\documentclass{PoS}

\usepackage{amsfonts,amssymb,amsmath,bm}
\usepackage{graphicx}

\title{Explicit construction of the pole part \\ of the three-gluon vertex.}

\ShortTitle{The pole part of the three-gluon vertex.}

\author{\speaker{David Ibanez}\\     
        Department of Theoretical Physics and IFIC,\\ 
        University of Valencia-CSIC,\\
        E-46100, Valencia, Spain.\\
        E-mail: \email{David.Gil@ific.uv.es}}

\abstract{
We present an explicit construction of the special part 
of the three gluon vertex, which  
incorporates the Schwinger mechanism into the 
Schwinger-Dyson equation of the gluon propagator, 
enabling the generation of a dynamical gluon mass. 
This vertex contains massless, longitudinally coupled poles,
acting effectively  as composite 
Nambu-Goldstone bosons, generated by the strong QCD dynamics.
The basic ingredients required for this construction 
are the longitudinal nature of this vertex
and the Slavnov-Taylor identities that it must satisfy, in order for 
gauge-invariance and BRST symmetry to remain intact in the 
presence of a gluon mass.}

\FullConference{International Workshop on QCD Green's Functions, Confinement 
and Phenomenology\\
		 5-9 September 2011\\
		 Trento, Italy}

\begin{document}

\section{Introduction}

One of the most crucial theoretical ingredients appearing 
in the analysis leading to the gauge-invariant 
generation of an effective    
gluon mass~\cite{Cornwall:1981zr,Bernard:1981pg,Aguilar:2008xm, Binosi:2009qm} 
is a special type of vertex, denoted by $V$, 
which contains massless, longitudinally coupled poles.
This vertex complements the all-order three-gluon vertex 
entering into the Schwinger-Dyson equations (SDEs) governing the gluon self-energy,
and is intimately connected with the famous Schwinger mechanism~\cite{Schwinger:1962tn}. 
The basic underlying assumption is that the strong QCD dynamics 
will lead to the formation of   
massless bound-state excitations, which, in turn, furnish   
the aforementioned poles that appear 
inside $V$~\cite{Jackiw:1973tr,Jackiw:1973ha,Cornwall:1973ts,Eichten:1974et,Poggio:1974qs}.  

Even though the presence of the vertex $V$ is indispensable for 
maintaining gauge invariance, its explicit closed form is 
yet undetermined~\cite{Aguilar:2011xe}. This is so, in part because, 
at the level of the ``one-loop dressed'' 
SDE analysis carried out so far, 
a great deal of information on the behavior of the gluon mass 
may be extracted  without explicit knowledge of the vertex $V$, 
invoking only some of its general 
properties, most notably the fact that it 
displays a completely longitudinal Lorentz structure, and that it satisfies very 
powerful  Slavnov-Taylor identities (STIs) and Ward identities (WIs)~\cite{Aguilar:2011ux}.

However, in order to be able to go beyond the ``one-loop dressed'' approximation 
in the SDE studies, the closed form of $V$ is absolutely necessary.
This necessity becomes particularly transparent within the 
formalism that has emerged from the synthesis between the pinch technique (PT)~\cite{Cornwall:1981zr,Cornwall:1989gv,Pilaftsis:1996fh, Binosi:2009qm,Aguilar:2006gr,Binosi:2007pi}
and the background-field method (BFM)~\cite{Abbott:1980hw}, known in the literature as 
the PT-BFM scheme~\cite{Aguilar:2008xm, Binosi:2009qm} . 
In the present work we carry out the explicit construction of the vertex $V$ 
within this particular formalism.

\section{General considerations.}

In this section we introduce the  appropriate  notation and
conventions, as well as the basic ingredients  that we will use in order
to construct  the pole part  of the three-gluon  vertex as well  as to
motivate the necessity of determine its explicitly form. Consider then
the full gluon propagator in the renormalizable $R_\xi$ gauges defined
as
\begin{equation} \label{gluon propagator}
\Delta_{\mu\nu}(q)=-i\bigg[P_{\mu\nu}(q)\Delta(q^2)+\xi\frac{q_\mu q_\nu}{q^4}\bigg],
\end{equation}
where
\begin{equation} \label{transverse projector}
P_{\mu\nu}(q)=g_{\mu\nu}-\frac{q^\mu q^\nu}{q^2}\,,
\end{equation}
is the dimensionless transverse projector, and $\xi$ the gauge fixing 
parameter (a color factor $\delta^{ab}$ has been factored out). 
The form factor $\Delta(q^2)$ is related to the all-order 
gluon self-energy $\Pi_{\mu\nu}(q)=P_{\mu\nu}(q)\Pi(q^2)$ through
\begin{equation} \label{inverse propagator}
\Delta^{-1}(q^2)=q^2+i\Pi(q^2)=q^2 J(q^2),
\end{equation}
where $J(q^2)$ is the {\it inverse} of the gluon dressing function. 
As a direct consequence of the gauge invariance of the theory, which after the 
gauge-fixing is encoded into the BRST symmetry, we know that 
the gluon self-energy is transverse, 
\begin{equation} \label{transversality}
q^\mu \Pi_{\mu\nu}(q)=0\,,
\end{equation}
to all orders in perturbation theory, as well as non-perturbatively, 
at the level of the SDE.

\begin{figure}[!t]
\begin{center}
\center{\includegraphics[scale=0.9]{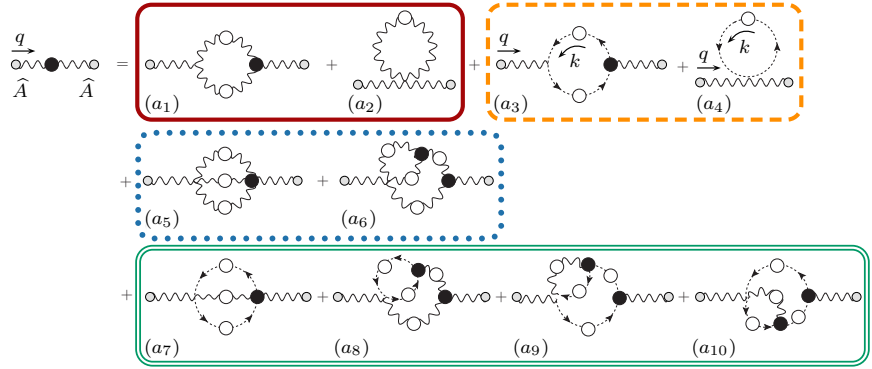}}
\caption{The SDE corresponding to the PT-BFM gluon self-energy $\Pi_{\mu\nu}$. 
The graphs inside each box form a gauge invariant subgroup, 
furnishing an individually transverse contribution. White (black) blobs 
denote full propagators (vertices).
 External background legs are indicated by the small gray circles.}
\label{fig1}
\end{center}
\end{figure}

As is well known, 
in the PT-BFM scheme, the SDE 
of the gluon propagator, shown in Fig.~\ref{fig1}, assumes the form~\cite{Aguilar:2008xm} ,
%
\begin{equation} \label{SDE gluon propagator}
\Delta^{-1}(q^2) P_{\mu\nu}(q) = \frac{q^2P_{\mu\nu}(q)+i\Pi_{\mu\nu}(q)}{[1+G(q^2)]^2} \,,
\end{equation}
where the function $G(q^2)$ 
is the $g_{\mu\nu}$ form factor in the Lorentz decomposition of the auxiliary function
\begin{eqnarray}\label{two-point auxiliary}
\Lambda_{\mu\nu}(q) &=& -ig^2C_A\int_k D(q-k)\Delta_\mu^\sigma(k)H_{\nu\sigma}(-q,q-k,k) \nonumber \\
&=& g_{\mu\nu}G(q^2)+\frac{q_\mu q_\nu}{q^2}L(q^2),
\end{eqnarray}
where $C_A$ is the Casimir eigenvalue of the adjoint representation of the gauge group, 
and $H_{\mu\nu}$ is the standard ghost-gluon kernel, shown diagrammatically 
in  Fig.~\ref{fig3}, together with  
the dressed-loop expansion of $\Lambda_{\mu\nu}$.

One of the most powerful properties of the PT-BFM formulation is that 
the transversality of the gluon self-energy is realized 
``blockwise''~\cite{Aguilar:2006gr}, 
following the pattern shown in Fig.\ref{fig1}.

\begin{figure}[!t]
\begin{center}
\center{\includegraphics[scale=0.5]{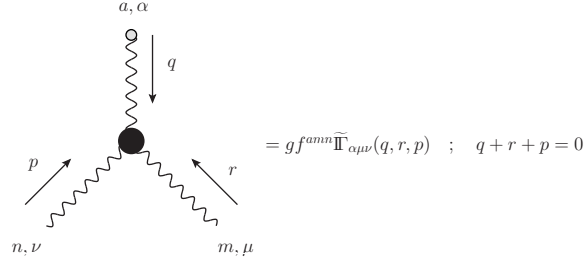}}
\caption{The BQQ vertex with the conventions for the momenta, color and Lorentz indices.}
\label{fig2}
\end{center}
\end{figure}

If we focus our attention on the "one-loop dressed" gluon contributions
to  the PT-BFM  gluon  self-energy,  given by  the  subset of  diagrams
$(a_1)$ and $(a_2)$, the relevant  Green's function to consider is the
three-gluon  vertex  with one  background  leg  and  two quantum  legs
denoted by  BQQ (see  Fig.~\ref{fig2}). This special  vertex satisfies
a WI when  contracted with  the momentum
$q_\alpha$ of  the background gluon leg,  and two
STIs when  contracted with the  momentum $r_\mu$ or $p_\nu$  of the
quantum gluon legs~\cite{Binosi:2011wi}, namely 
\begin{eqnarray} \label{BQQ STI}
&& q^\alpha \widetilde{\mathrm{I}\!\Gamma}_{\alpha\mu\nu}(q,r,p) = p^2J(p^2) P_{\mu\nu}(p) - r^2J(r^2) P_{\mu\nu}(r), \nonumber\\
&& r^\mu \widetilde{\mathrm{I}\!\Gamma}_{\alpha\mu\nu}(q,r,p) = F(r^2)\left[q^2\widetilde{J}(q^2)P_\alpha^\mu(q)H_{\mu\nu}(q,r,p) - p^2J(p^2) P_\nu^\mu(p)\widetilde{H}_{\mu\alpha}(p,r,q)\right], \nonumber\\
&& p^\nu \widetilde{\mathrm{I}\!\Gamma}_{\alpha\mu\nu}(q,r,p) = F(p^2)\left[r^2J(r^2) P_\mu^\nu(r)\widetilde{H}_{\nu\alpha}(r,p,q) - q^2\widetilde{J}(q^2)P_\alpha^\nu(q)H_{\nu\mu}(q,p,r)\right].
\end{eqnarray}
In these identities the ghost-gluon kernel $\widetilde{H}_{\mu\nu}$ is obtained 
from the conventional  ${H}_{\mu\nu}$
by replacing the external gluon by a background gluon, as shown in 
Fig.~\ref{fig3}.
The quantity $F(q^2)$ represents 
the ghost dressing function, related to the ghost propagator $D(q^2)$ through
\begin{equation} \label{dressing ghost}
D(q^2)=\frac{F(q^2)}{q^2}.
\end{equation}
\begin{figure}[!b]
\center{\includegraphics[scale=0.45]{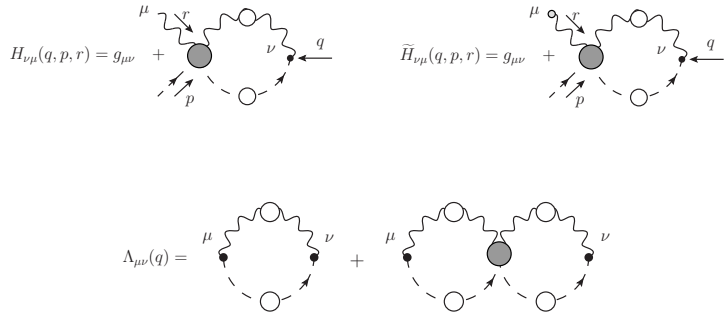}}
\caption{Diagrammatic representation of the auxiliary functions $H$, $\widetilde{H}$ and $\Lambda$. 
White blobs represent dressed propagators, while gray blobs denote one-particle irreducible kernels with respect to vertical cuts.}
\label{fig3}
\end{figure}
Finally, the function $\widetilde{J}(q^2)$ corresponds to the inverse dressing function 
of the mixed ``background-quantum'' gluon propagator (one background and 
one quantum gluons entering, BQ), denoted by $\widetilde{\Delta}(q^2)$. 
This latter propagator, together with the conventional gluon propagator (two quantum gluons entering, QQ), 
denoted by $\Delta(q^2)$,  
and the background gluon propagator (two background gluons entering, BB),
denoted by $\widehat{\Delta}(q^2)$, 
are the three types of gluon propagators that appear naturally in the BFM formalism. 
They are related
by the so called ``background-quantum identities'' (BQIs)~\cite{Grassi:1999tp, Binosi:2002ez}
\begin{eqnarray} \label{BQI's}
\Delta(q^2) &=& [1 + G(q^2)]^2 \widehat{\Delta}(q^2),\nonumber\\
\Delta(q^2) &=& [1 + G(q^2)] \widetilde{\Delta}(q^2),\nonumber\\
\widetilde{\Delta}(q^2) &=& [1 + G(q^2)] \widehat{\Delta}(q^2).
\end{eqnarray}

Now, if we want to trigger the Schwinger mechanism, 
a pole vertex $\widetilde{V}_{\alpha\mu\nu}(q,r,p)$ containing longitudinally 
coupled massless bound-state excitations must be added to the conventional (fully-dressed) 
BQQ three-gluon vertex $\widetilde{\mathrm{I}\!\Gamma}_{\alpha\mu\nu}(q,r,p)$, 
giving rise to the new full vertex $\widetilde{\mathrm{I}\!\Gamma}'_{\alpha\mu\nu}(q,r,p)$ defined as~\cite{Aguilar:2011xe}
\begin{equation}\label{fullvertex}
\widetilde{\mathrm{I}\!\Gamma}'_{\alpha\mu\nu}(q,r,p) = \widetilde{\mathrm{I}\!\Gamma}_{\alpha\mu\nu}(q,r,p) + \widetilde{V}_{\alpha\mu\nu}(q,r,p).
\end{equation}
The presence of this pole vertex enforces the gauge-invariance of the 
theory in the presence of masses. Specifically, when  
the gluon propagator becomes effectively massive, 
assuming the form~\cite{Aguilar:2011ux,Cucchieri:2007md,Oliveira:2008uf}
\begin{equation} \label{massive propagator}
\Delta^{-1}_m(q^2) = q^2J(q^2)-m^2(q^2),
\end{equation}
the full vertex  $\widetilde{\mathrm{I}\!\Gamma}'$ ought to preserve the
fundamental  property  (\ref{transversality});  so, it must satisfy  the  same
formal    STI's   (\ref{BQQ   STI}),    but   with    the   replacement
$\Delta^{-1}\rightarrow\Delta^{-1}_m$.   This   requirement  will   be
automatically   fulfilled  if   we  demand   that  the   pole  vertex
$\widetilde{V}$ satisfies the following STI's~\cite{Aguilar:2011xe},
\begin{eqnarray} \label{pole BQQ STI}
&& q^\alpha \widetilde{V}_{\alpha\mu\nu}(q,r,p) = m^2(r^2) P_{\mu\nu}(r) - m^2(p^2) P_{\mu\nu}(p), \nonumber\\
&& r^\mu \widetilde{V}_{\alpha\mu\nu}(q,r,p) = F(r^2)\left[m^2(p^2) P_\nu^\mu(p)\widetilde{H}_{\mu\alpha}(p,r,q)-\widetilde{m}^2(q^2)P_\alpha^\mu(q)H_{\mu\nu}(q,r,p)\right], \nonumber\\
&& p^\nu \widetilde{V}_{\alpha\mu\nu}(q,r,p) = F(p^2)\left[\widetilde{m}^2(q^2)P_\alpha^\nu(q)H_{\nu\mu}(q,p,r)-m^2(r^2) P_\mu^\nu(r)\widetilde{H}_{\nu\alpha}(r,p,q)\right].
\end{eqnarray}
The mass $\widetilde{m}$ appearing in Eq.~(\ref{pole BQQ STI}) denotes the mass  
of the mixed background-quantum gluon propagator $\widetilde{\Delta}(q^2)$, and it is known to satisfy 
the same BQI as the full gluon propagator, namely Eq.(\ref{BQI's}), i.e~\cite{Aguilar:2011ux}
\begin{equation} \label{mass BQI}
\widetilde{m}^2(q^2) = [1 + G(q^2)] m^2(q^2). 
\end{equation}

Finally, observe that the  "two-loop dressed" gluon contribution to the
PT-BFM gluon self-energy, given by  the subset of diagrams $(a_5)$ and
$(a_6)$  in Fig.~\ref{fig1}, contains  an internal  three-gluon vertex
with three  quantum gluon  legs (QQQ), as  well as a  four-gluon vertex
with one  background and  three quantum gluon  legs (BQQQ).  This BQQQ
vertex satisfies the following WI  when contracted with respect to the
background gluon leg~\cite{Aguilar:2006gr},
\begin{eqnarray} \label{BQQQ WI}
q_1^\mu \widetilde{\mathrm{I}\!\Gamma}^{abcd}_{\mu\nu\alpha\beta}(q_1,q_2,q_3,q_4) &=& igf^{abx}\mathrm{I}\!\Gamma^{cdx}_{\alpha\beta\nu}(q_3,q_4,q_1+q_2) \nonumber\\
&+& igf^{acx}\mathrm{I}\!\Gamma^{dbx}_{\beta\nu\alpha}(q_4,q_2,q_1+q_3) \nonumber\\
&+& igf^{adx}\mathrm{I}\!\Gamma^{bcx}_{\nu\alpha\beta}(q_2,q_3,q_1+q_4).
\end{eqnarray}
Therefore, the  description of the  "two-loop dressed" gluon  block in
the presence of vertices with pole structures requires the knowledge of
the pole QQQ  three-gluon vertex, denoted by $V$.  In this case, 
the background leg $q_\alpha$  becomes quantum, and the Abelian-like WI in
(\ref{pole BQQ STI}) is replaced  by an STI, namely
\begin{equation} 
\label{pole QQQ STI}
q^\alpha V_{\alpha\mu\nu}(q,r,p) = F(q^2)\left[m^2(r^2)P^\alpha_\mu(r)H_{\alpha\nu}(r,q,p)-m^2(p^2)P^\alpha_\nu(p)H_{\alpha\mu}(p,q,r)\right],
\end{equation}
while the STIs with respect to the other two legs are those of Eq.~(\ref{pole BQQ STI}), but with the 
``tilded'' quantities replaced by conventional ones.

\section{Explicit construction.}
Turns out the the explicit closed form of the two pole vertices in question, $\widetilde{V}$ and $V$,  
may be determined from the STIs they satisfy, and the requirement  
of complete  longitudinality, i.e,  
condition~\cite{Aguilar:2011xe}
\begin{equation}\label{longitudinally}
P^{\alpha\beta}(q) P^{\mu\rho}(r) P^{\nu\sigma}(p) \widetilde{V}_{\beta\rho\sigma}(q,r,p) = 0.
\end{equation}
Specifically, opening up transverse projectors in (\ref{longitudinally}), 
one can write the entire vertex in terms of its own divergences,
\begin{eqnarray} \label{divergences}
\widetilde{V}_{\alpha\mu\nu}(q,r,p) &=& \frac{q_\alpha}{q^2}q^\beta\widetilde{V}_{\beta\mu\nu} + \frac{r_\mu}{r^2}r^\rho\widetilde{V}_{\alpha\rho\nu} + \frac{p_\nu}{p^2}p^\sigma\widetilde{V}_{\alpha\mu\sigma} - \frac{q_\alpha r_\mu}{q^2r^2}q^\beta r^\rho \widetilde{V}_{\beta\rho\nu} - \frac{q_\alpha p_\nu}{q^2p^2}q^\beta p^\sigma \widetilde{V}_{\beta\mu\sigma} \nonumber \\
&-& \frac{r_\mu p_\nu}{r^2p^2}r^\rho p^\sigma \widetilde{V}_{\alpha\rho\sigma} + \frac{q_\alpha r_\mu p_\nu}{q^2r^2p^2}q^\beta r^\rho p^\sigma \widetilde{V}_{\beta\rho\sigma}.
\end{eqnarray}
Note that the last term will not contribute because if we apply the STI's,
\begin{equation}
q^\beta r^\rho p^\sigma \widetilde{V}_{\beta\rho\sigma}(q,r,p) = 0.
\end{equation}
So, using (\ref{pole BQQ STI}) to evaluate the various terms,  and after a straightforward rearrangement,  
we obtain the following expression for the pole part of the BQQ vertex,
\begin{eqnarray} \label{poleBQQ}
\widetilde{V}_{\alpha\mu\nu}(q,r,p) &=& \frac{q_\alpha}{q^2}\left[m^2(r^2)-m^2(p^2)\right]P_\mu^\rho(r)P_{\rho\nu}(p) \nonumber\\
&+& D(r^2)\left[m^2(p^2)P_\nu^\rho(p)\widetilde{H}_{\rho\alpha}(p,r,q)-\widetilde{m}^2(q^2)P_\alpha^\rho(q)P_\nu^\sigma(p)H_{\rho\sigma}(q,r,p)\right]r_\mu \nonumber\\
&+& D(p^2)\left[\widetilde{m}^2(q^2)P_\alpha^\rho(q)H_{\rho\mu}(q,p,r)-m^2(r^2)P_\mu^\rho(r)\widetilde{H}_{\rho\alpha}(r,p,q)\right]p_\nu .
\end{eqnarray}

Applying the same procedure but using now the STIs (\ref{pole QQQ STI}) 
as well as the longitudinally coupled condition (\ref{longitudinally}), 
we derive the closed expression for the pole part of the QQQ vertex
\begin{eqnarray} \label{poleQQQ}
V_{\alpha\mu\nu}(q,r,p) &=& D(q^2)\left[m^2(r^2)H_{\rho\sigma}(r,q,p)-m^2(p^2)H_{\sigma\rho}(p,q,r)\right]P_\mu^\rho(r)P_\nu^\sigma(p)q_\alpha \nonumber\\
&+& D(r^2)\left[m^2(p^2)P^\rho_\nu(p)H_{\rho\alpha}(p,r,q)-m^2(q^2)P_\alpha^\rho(q)P_\nu^\sigma(p)H_{\rho\sigma}(q,r,p)\right]r_\mu \nonumber\\
&+& D(p^2)\left[m^2(q^2)P^\rho_\alpha(q)H_{\rho\mu}(q,p,r)-m^2(r^2)P_\mu^\rho(r)H_{\rho\alpha}(r,p,q)\right]p_\nu .
\end{eqnarray}

Now  we need  to discuss  some points related to the  self-consistency  of our
vertex  construction.  Observe that  in  order  to obtain  expressions
(\ref{poleBQQ}) and (\ref{poleQQQ}) one must apply sequentially the WI
and  the STIs. In  doing so,  the Bose  symmetry of  both vertices
is no longer explicit, and the result obtained 
is not manifestly symmetric under the
quantum  gluon legs exchange.  Furthermore, seemingly different  expressions are
obtained, depending on  which of the two momenta acts first on $\widetilde{V}$. 
However, if one imposes the simple requirement of algebraic 
commutativity between  the  WI  and the  STIs  satisfied by  the
three-gluon vertex, the Bose symmetry becomes manifest.
For example,  using (\ref{poleBQQ}) we
can see that the elementary requirement 
\begin{equation}\label{commutativity tilde}
q^\alpha r^\mu \widetilde{V}_{\alpha\mu\nu}(q,r,p) = r^\mu q^\alpha \widetilde{V}_{\alpha\mu\nu}(q,r,p)\,,
\end{equation}
gives rise to the following identity 
\begin{equation}\label{constraint H tilde}
F(r^2)P_\nu^\mu(p) q^\alpha \widetilde{H}_{\mu\alpha}(p,r,q) = - r_\mu P_\nu^\mu(p).
\end{equation}
A similar identity is obtained by imposing the requirement of (\ref{commutativity tilde})
at the level of $V$, namely 
\begin{equation} \label{constraint H}
F(r^2)P_\nu^\mu(p) q^\alpha H_{\mu\alpha}(p,r,q) = - F(q^2) P_\nu^\mu(p) r^\alpha H_{\mu\alpha}(p,q,r).
\end{equation}

\begin{figure}[ht]
\center{\includegraphics[scale=0.66]{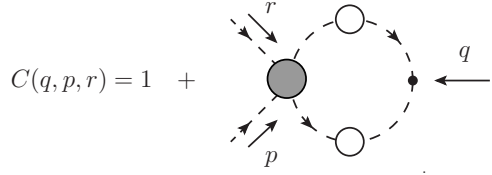}}
\caption{Diagrammatic representation of the auxiliary function $C(q,p,r)$.}
\label{fig4}
\end{figure}

Quite remarkably, the  identities (\ref{constraint  H
tilde}) and (\ref{constraint H}) are a direct consequence of 
WI and the
STI  that the kernels $H$ and $\widetilde{H}$ satisfy,  when  they  are
contracted with  the momentum of  the background or quantum  gluon leg, namely~\cite{Binosi:2011wi},
\begin{eqnarray} \label{ghostauxiliary STI}
&& q^\alpha \widetilde{H}_{\mu\alpha}(p,r,q) = -p_\mu F^{-1}(p^2) - r_\mu F^{-1}(r^2), \nonumber\\
&& q^\alpha H_{\mu\alpha}(p,r,q) = -F(q^2)\left[p_\mu F^{-1}(p^2)C(q,r,p) + r^\alpha F^{-1}(r^2) H_{\mu\alpha}(p,q,r)\right],
\end{eqnarray}
where $C(q,r,p)$ is the auxiliary function that characterizes 
the four-ghost kernel (see Fig.~\ref{fig4}). 
Indeed, use of (\ref{ghostauxiliary STI}) into (\ref{constraint  H tilde}) 
and (\ref{constraint H}), respectively, leads to a trivial identity. 
Conversely, one may actually derive (\ref{ghostauxiliary STI}) from (\ref{constraint  H
tilde}) and (\ref{constraint H});   
for example, starting with (\ref{constraint H tilde}), and using also the
identities~\cite{Binosi:2011wi}
\begin{eqnarray}
p^\mu \widetilde{H}_{\mu\alpha}(p,r,q) &=& r_\alpha F^{-1}(r^2)-\widetilde{\Gamma}_\alpha(r,q,p), \nonumber\\
q^\alpha \widetilde{\Gamma}_\alpha(r,q,p) &=& p^2 F^{-1}(p^2) - r^2 F^{-1}(r^2),
\end{eqnarray}
one can easily reproduce (\ref{ghostauxiliary STI})

Evidently, these constraints allow us to cast the pole part of the BQQ vertex 
into a manifestly Bose symmetric form with respect to the quantum legs, 
\begin{equation}\label{BoseBQQ}
\widetilde{V}_{\alpha\mu\nu}(q,r,p) = \frac{q_\alpha}{q^2}\left[m^2(r^2)-m^2(p^2)\right]P_\mu^\rho(r)P_{\rho\nu}(p) + \widetilde{I}_{\alpha\mu\nu}(q,r,p) - \widetilde{I}_{\alpha\nu\mu}(q,p,r),
\end{equation}
with
\begin{eqnarray}
\widetilde{I}_{\alpha\mu\nu}(q,r,p) &=& D(r^2)m^2(p^2)P_\nu^\rho(p)\widetilde{H}_{\rho\alpha}(p,r,q)r_\mu \nonumber\\ 
&-& \frac{r_\mu}{2}D(r^2)\widetilde{m}^2(q^2)P_\alpha^\rho(q)\left[g_\nu^\sigma + P_\nu^\sigma(p)\right] H_{\rho\sigma}(q,r,p).
\end{eqnarray}
Finally, for the pole part of the QQQ vertex, the Bose symmetric expression reads
\begin{equation}\label{BoseQQQ}
V_{\alpha\mu\nu}(q,r,p) = I_{\alpha\mu\nu}(q,r,p) - I_{\mu\alpha\nu}(r,q,p) - I_{\alpha\nu\mu}(q,p,r),
\end{equation}
with
\begin{eqnarray}
I_{\alpha\mu\nu}(q,r,p) &=& \frac{q_\alpha}{2}D(q^2)\left[m^2(r^2)P_\mu^\rho(r)H_{\rho\nu}(r,q,p)-m^2(p^2)P_\nu^\rho(p)H_{\rho\mu}(p,q,r)\right] \nonumber\\
&+& \frac{q_\alpha}{2}D(q^2)\left[m^2(r^2)H_{\rho\sigma}(r,q,p)-m^2(p^2)H_{\sigma\rho}(p,q,r)\right]P_\mu^\rho(r)P_\nu^\sigma(p).
\end{eqnarray}

\section{Conclusions.}

In this work we have reported the explicit closed form of the pole parts of two particular vertices, 
which are intimately connected to the phenomenon of gluon mass generation, as described 
within the PT-BFM  formalism. 
Specifically, we have 
determined the pole parts of the BQQ and QQQ vertices, denoted by ${\tilde V}$ and $V$, respectively. 
The only ingredient necessary for this construction is the longitudinal nature of ${\tilde V}$ and $V$
and the STIs and WIs that they must satisfy. These two vertices are expected to form an integral 
part of the ongoing SDE studies that aim to determine the precise quantitative details 
of the gluon mass generation mechanism.

\acknowledgments

This research is supported by the European FEDER and Spanish MICINN under grant FPA2008-02878.

\end{document}